\documentclass{ifacconf}

\usepackage{graphicx}
\usepackage{natbib}        
\usepackage{xcolor}
\usepackage{amsmath}
\usepackage{amssymb}
\usepackage{algorithm}
\usepackage{diagbox}
\usepackage{wasysym}
\usepackage{enumitem}
\usepackage{fdsymbol}
\usepackage{comment}
\usepackage{soul}
\DeclareMathOperator*{\argmin}{arg\,min}

\newtheorem{lemma}{Lemma}
\newtheorem{remark}{Remark}
\newtheorem{assumption}{Assumption}
\begin{document}
\begin{frontmatter}

\title{
Precision Data-enabled Koopman-type Inverse  Operators
for  Linear Systems  
\thanksref{footnoteinfo}} 

\thanks[footnoteinfo]{This work was supported by NSF Grant CMMI 1824660.}

\author[First]{Leon (Liangwu) Yan} 
\author[Second]{Santosh Devasia} 

\address[First]{Mechanical Engineering Department, University of Washington, Seattle, WA 98195-2600, USA (e-mail: liangy00@uw.edu).}
\address[Second]{Mechanical Engineering Department, University of Washington, Seattle, WA 98195-2600, USA (e-mail: sdevasia@uw.edu)}

\begin{abstract}
The advent of easy access to large amount of data has sparked interest in directly developing the relationships between input and output of dynamic systems. 
A challenge is that  in addition to the applied input and the measured output, the dynamics  can also depend on hidden states that are not directly measured.
The main contribution of this work is 
to identify the information needed (in particular, the past history of the output) 
to 
remove the hidden state dependence in Koopman-type inverse operators for linear systems. 
Additionally, it is shown that the time history of the output  should be augmented with the instantaneous time derivatives of the output to achieve precision of the inverse operator. 
This insight into the required output (history and instantaneous derivative) information, 
to remove the hidden-state dependence and improve  the precision of data-enabled  inverse operators, is illustrated with an example system.
\end{abstract}

\begin{keyword}
Dynamics, hidden state dependency, neural-network models, inverse models
\end{keyword}

\end{frontmatter}

\section{Introduction}
With increasing ease of collecting data and low cost storage, there is increasing interest to use data-enabled methods developing models for prediction and   control,~\cite{abraham2017model,mamakoukas2021derivative,hewing2020learning, asadi2021gaussian,piche2000neural,kabzan2019learning,kocijan2004gaussian}. 
Such models can be optimized to best fit the data and methods are also available to  estimate the error bounds on the predictions, e.g., using  predicted time derivatives of the observables~\cite{mamakoukas2021derivative}.
However, the conditions under which such data-enabled models can achieve sufficient precision remains unclear.
A major challenge is that the model (i.e., the relationship between the input and the measurable outputs) can be dependent on the system's internal states, which are  hidden in the sense that they are not directly measured, nor  inferred using standard observer designs since they require prior knowledge of the system dynamics. 

Several approaches are available to address the lack of direct access to the hidden states. 
One approach is to represent the  dynamics through Markov models with a predefined number of hidden states, and then minimize the model prediction error~\cite{tarbouriech2020active,yoon2019hidden,pohle2017selecting}. A difficulty is that the optimal selection of the number of hidden states can be computationally expensive, and there is no guarantee that the resulting models will achieve the desired precision. A second class of approaches to handle the lack of direct access to the hidden states is to model the system dynamics (flow) 
in a lifted observable space (with generalized functions of the observables) using  Koopman operator theory \cite{schmid2008dynamic,mezic2005spectral}. Recent techniques  include  sparse identification of nonlinear dynamical systems (SINDy)~\cite{brunton2016discovering} and linearization Dynamic Mode Decomposition(DMD)~\cite{kutz2016dynamic}. Nevertheless, with a finite number of states,  there is uncertainty about how to  select  a sufficient set of  generalized observable functions to achieve a specified level of prediction precision. 
A third class of approaches is to use time history of the input and output  data  to find forward models, e.g., with  (i)~transfer function models in the frequency domain~\cite{devasia2017iterative,yan2021mimo};  (ii)~autoregressive models with extra input (ARX)~\cite{ljung1987theory} as well as  nonlinear ARX (NARX)~\cite{kocijan2004gaussian,pham2010hybrid}; (iii)~time-delayed information in the Koopman operator framework~\cite{kamb2020time}; and fitting a relation between the time-delayed output data 
and the inverse input \cite{butterworth2012analysis,blanken2020kernel,aarnoudse2021control}. 
Again, determining the type of  data needed
to capture the input-output relationship (with high  precision) when models are not available a-priori remains uncertain. When precision of the inverse is not sufficient, it can be improved using  iterative techniques, with the  inverse of the plant considered as the learning operator, \cite{ghosh2001iterative,fine2009model,teng2015comparison,spiegel2021iterative}. Nevertheless, increasing the precision of the inverse model can improve ILC convergence.

The goal of this article is to identify the type of output data needed to develop  inverse (output-to-input) operators, with a desired level of  precision.
Rather than the two step processes of first  learning forward models and second using  model-predictive control (MPC) to optimally select the control input, the proposed approach seeks to solve the inverse problem of directly finding the input for a given output, e.g., similar to~\cite{dev96ac,willems2005note}. 
In particular,  the relative degree of the system is used to identify the number of time derivatives that need to be added to input-output data  to facilitate precision data-enabled learning of the inverse operator.

Previous works on inversion of system dynamics, using known models of the system, have shown that  the  impact of neglecting the boundary conditions of the internal states can be made arbitrarily small~\cite{zou1999preview,zou2007precision} by choosing a sufficiently large time history  of the desired output and its derivatives.
This motivates the proposed data-enabled algorithm to learn the inverse operator directly from input-output data (without the need to explicitly capture the hidden state dynamics) by using time-delayed observations of the  output, along with the output's time derivatives.  

The main contribution of this paper is to  propose a Koopman-type
time-delay and output-derivative-based data-enabled 
inverse operator that minimizes the impact of  the hidden state dependency and achieves precision (illustrated with a simulation example). Overall, the work provides insight into the need for including derivative features and time history to achieve precision in Koopman-type inverse operators. Even for forward Koopman-type operators (which only depend on past observable outputs) it is shown that that the output-derivative at the current time instant needs to be included for precision prediction.

\section{Problem formulation and solution}
The inverse operator
is developed for 
linear time-invariant (LTI) single-input-single-output (SISO) system. 
Let the system be
\begin{align}
\Dot{x}(t)&=Ax(t)+Bu(t)\label{eq:X_dynamics}\\
y(t)&=Cx(t)\label{eq:output}
\end{align}
with states $x(t)\in \mathbb{R}^{n}$, input $u(t)\in \mathbb{R}$ and output $y(t)\in \mathbb{R}$ with  matrices $A \in \mathbb{R}^n \times  \mathbb{R}^n, B\in \mathbb{R}^n \times 1,C \in 1\times \mathbb{R}^n$.

\begin{assumption}[System properties]
The system described in (\ref{eq:X_dynamics}) and (\ref{eq:output}) is stable (i.e., $A$ is Hurwitz), hyperbolic (no zeros on the imaginary axis), and has relative degree $r \le n$ (i.e., the difference between the number of poles and the number of zeros).
\label{assum:relative_degree}
\end{assumption}
\begin{assumption}
The desired output $y_d$, specified in inverse 
operator
problems,  is sufficiently  smooth, and has  bounded time derivatives upto the relative degree $r$.
\end{assumption}

\subsection{Hidden state dependency}
The system state $x$ can split into state components $\xi$ that directly depend on the output and its time derivatives
\begin{align}
\xi(t) &= \begin{bmatrix}y(t),\Dot{y}(t),\dots,\frac{d^{r-1} y(t)}{d t^{r-1}}\end{bmatrix}'\in \mathbb{R}^{r\times 1}
\label{eq:xi_def}
\end{align}
and internal states $\eta$, 
\begin{equation}
\begin{bmatrix}
\xi(t)\\ \eta(t)
\end{bmatrix}=Sx(t)
\label{eq:coord_trans}
\end{equation}
such that in the new coordinates, (\ref{eq:X_dynamics}) can be written as, e.g., see \cite{marino1995nonlinear}, Example 4.1.3, 
\begin{align}
\Dot{\xi}(t)&=A_1\xi(t)+A_2\eta(t)+B_1u(t) \label{eq:xi_dynamic}\\
\Dot{\eta}(t)&=A_3y(t)+A_4\eta(t) \label{eq:eta_dynamic}
\end{align}
where
\begin{equation*}
B_1= 
\begin{bmatrix}
0\\ 0 \\\vdots\\b_{n-r}
\end{bmatrix}\in\mathbb{R}^{r\times 1},\quad 
A_3=\begin{bmatrix}
0\\0\\\vdots \\ 1/b_{n-r}
\end{bmatrix}, 
\end{equation*}
\begin{equation*}
A_4=\begin{bmatrix}
0&1&\dots& 0\\
\vdots&\vdots&\ddots&\vdots\\
0&0&\dots&1\\
-b_0/b_{n-r}&-b_1/b_{n-r}&\dots&-b_{n-r-1} / b_{n-r},
\end{bmatrix}    
\end{equation*}
and the eigenvalues of matrix $A_4$ are the zeros of the transfer function of  system (\ref{eq:X_dynamics}) and (\ref{eq:output}).
\begin{equation}
G(s) = \frac{Y(s)}{U(s)}=
\frac{b_0+b_1s+\dots+b_{n-r}s^{n-r}}{a_0+a_1s+\dots+a_{n-1}s^{n-1}+s^n}.
\label{eq:transfer_func}
\end{equation}
Note that the internal state $\eta$ is only driven by the output $y = \xi_1$.
Moreover, due to the relative degree $r$ assumption,  the input $u$ is directly related to the  $r^{th}$  derivative of the output, and therefore, the $r^{th}$  row of (\ref{eq:xi_dynamic}) can be written as 
\begin{equation}
\begin{split}
y^{(r)}(t)\triangleq \frac{d^{r}y(t)}{dt^{r}}&=C A^rx +CA^{r-1}Bu(t)\\
&=CA^r S^{-1}\begin{bmatrix}
\xi(t)\\\eta(t)
\end{bmatrix}+b_{n-r}u(t) \\
&= A_{\xi} \xi(t) +A_{\eta} \eta(t)  + b_{n-r}u(t),
\end{split}
\label{eq:rel_degree_connection}
\end{equation}
and the matrices $A_1$ and $A_2$ in (\ref{eq:xi_dynamic}) are given by 
\begin{equation*}
A_1=\begin{bmatrix}
\begin{matrix} 
0&1&\dots& 0\\
\vdots&\vdots&\ddots&\vdots\\
0&0&\dots&1
\end{matrix} \\
\hline \\[-0.1in]
A_{\xi}
\end{bmatrix}, ~~
A_2=\begin{bmatrix}
\begin{matrix} 
0&0&\dots& 0\\
\vdots&\vdots&\ddots&\vdots\\
0&0&\dots&0
\end{matrix} \\
\hline \\[-0.1in]
A_{\eta} 
\end{bmatrix}.
\end{equation*}
where $A_{\xi}$ and $A_{\eta}$ are the last rows of matrices $A_1$ and $A_2$ respectively.

\subsection{Research problem}

\vspace{-0.1in}
The desired output and its derivatives, $(y_d^{(r)}, \xi_d)$ can be used to predict the inverse input $u_d$ from (\ref{eq:rel_degree_connection}) , as
\begin{equation}
u_d(t)  = b_{n-r}^{-1}\left[ 
y_d^{(r)}(t)- A_{\xi} \xi_d(t) -
A_{\eta} \eta_d(t)
\right], 
\label{eq:inv_model}
\end{equation}
which depends on the internal states $\eta$ that are hidden or not directly measured. The goal is to minimize the hidden state effects on the 
inverse model, by addressing the following research problems.
\begin{enumerate}[label=(\roman*)]
\item Finding the hidden state from output: Develop an operator that maps the time history of the output $y$ with length $T$ to an estimate of the hidden state $\eta$ at time $t$
\begin{equation}
\hat\eta(t) = \hat{\mathbb{H}}[y(t-T:t)].
\label{op_internal_eta}
\end{equation}
\item Koopman-type inverse operator: Using the operator
in (\ref{op_internal_eta}), develop a data-enabled
Koopman-type inverse operator $\hat{\mathbb{G}}^{-1}$ 
that uses the history of the desired output and its time derivatives to predict the inverse input as 
\begin{align}
\hat{u}_d(t) &= \hat{\mathbb{G}}^{-1}[y_d(t-T:t),\xi_d(t),y^{(r)}_d(t)]\label{op_inverse_min_phase}.
\end{align}
\item 
Inverse operator
precision: Quantify the error $\|\hat{u}_d(t)-u_d(t)\|_2$
dependence on each argument of $\hat{\mathbb{G}}^{-1}$.
\end{enumerate} 

\subsection{Solution}

\subsubsection{Finding the hidden state from output}
If the system is minimum-phase ($A_4$ is Hurwitz), i.e., (\ref{eq:transfer_func}) has no zeros on the right half plane, then $\eta(t)$ can be obtained from the history of the output by solving (\ref{eq:eta_dynamic})
\begin{equation}
\begin{split}
\eta(t) 
&= \int_{-\infty}^t e^{A_4(t-\tau)}A_3y(\tau)d\tau\\
&\triangleq \mathbb{H}[y(-\infty:t)].
\end{split}
\label{eq:unknown_state_by_integral}
\end{equation}
In practice, such an operator is hard to capture in a data-enabled way since it requires an infinite window. Therefore, an estimate $\hat{\eta}$ is obtained with an approximate operator $\hat{\mathbb{H}}$ with a 
 finite time history 
length $T$ is defined

\begin{equation}
\begin{split}
\hat{\eta}(t)
&\triangleq\int_{t-T}^t e^{A_4(t-\tau)}A_3y(\tau)d\tau\\
&\triangleq \hat{\mathbb{H}}[y(t-T:t)].
\end{split}
\label{eq_approx_unknown_state}
\end{equation}
The approximate operator $\hat{\mathbb{H}}$  approaches the exact operator ${\mathbb{H}}$ exponentially as the 
time history
$T$ increases.

\begin{lemma}
\label{lemma_internal_state_estimate}
If the  output trajectory
is bounded, 
\begin{equation}
    M=\max_{\tau\in[-\infty, t-T]}\|y(\tau)\|_2 < \infty, 
    \label{eq_output_bound}
\end{equation}
then the error in computing the hidden state $\eta(t)$ decays exponentially with the
time history
$T$, i.e., there exists positive scalars $\alpha_1>0, \beta_1>0$ such that 
\begin{equation}
\begin{split}
\|\Delta \eta(t)\|_2\triangleq\|\eta(t)-\hat{\eta}(t)\|_{2}
\le \beta_1 e^{-\alpha_1 T}.
\end{split}
\label{eq_eta_err_bound_exp}
\end{equation}
\end{lemma}

\begin{pf}
Since the system is assumed to be minimum phase, the 
and the eigenvalues of matrix $A_4$, which are the zeros of the transfer function of  system (\ref{eq:X_dynamics}), lie in the open left-half of the complex plane, i.e., the matrix
$A_4$ is Hurwitz.
Then, 
there exists positive scalars $\kappa_1 >0, \alpha_1 >0$ such that,~\cite{desoer1975feedback} 
\begin{equation}
\|e^{A_4t}\|_{2}\le \kappa_1 e^{-\alpha_1 t}.
\label{eq:exponeital_decay}
\end{equation}
Then, from (\ref{eq:unknown_state_by_integral},\ref{eq_approx_unknown_state}), the approximation error can be bounded as

\begin{equation}
\begin{split}
\|\eta(t)-\hat{\eta}(t)\|_{2}&=\left\|
\int_{-\infty}^{t-T}e^{A_4(t-\tau)}A_3y(\tau)d\tau
\right\|_{2}\\
&\le M\|A_3\|_{2}
\int_{-\infty}^{t-T}\kappa_1 e^{-\alpha_1 (t-\tau)} d\tau \\
& \qquad {\mbox{using (\ref{eq_output_bound}, \ref{eq:exponeital_decay})} } \\
&= M\|A_3\|_{2}
\int_{T}^{+\infty}\kappa_1 e^{-\alpha_1 \tau'} d\tau'\\
&=M\|A_3\|_{2}
\frac{\kappa_1}{\alpha_1}e^{-\alpha_1 T}.
\end{split}    
\end{equation}
The result follows with 
\begin{equation}
    \beta_1 = 
M\|A_3\|_{2}
\frac{\kappa_1}{\alpha_1} .
    \label{eq_output_bound_2}
\end{equation}
\end{pf}

\subsubsection{Koopman-type inverse operator}
Given an estimate $\hat{\eta}$ of the internal state $\eta $, the inverse 
operator
prediction in (\ref{op_inverse_min_phase}) can  be estimated as
\begin{align}
\hat{u}_d(t)
&=b_{n-r}^{-1}\left[ 
y_d^{(r)}(t)- A_{\xi} \xi_d(t) -A_{\eta} \hat{\eta}_d(t)
\right] \nonumber \\
& = 
b_{n-r}^{-1}\left[ 
y_d^{(r)}(t)- A_{\xi} \xi_d(t) -A_{\eta} \hat{\mathbb{H}}[y_d(t-T:t)]
\right] \nonumber \\
& \qquad {\mbox{using (\ref{eq_approx_unknown_state})} } \nonumber \\
&\triangleq \hat{\mathbb{G}}^{-1}[y^{(r)}_d(t), \xi_d(t), y_d(t-T:t)].
\label{eq:inv_op_derivation}
\end{align}

\begin{figure}[!ht]
\centering
\includegraphics[width=0.95\columnwidth]{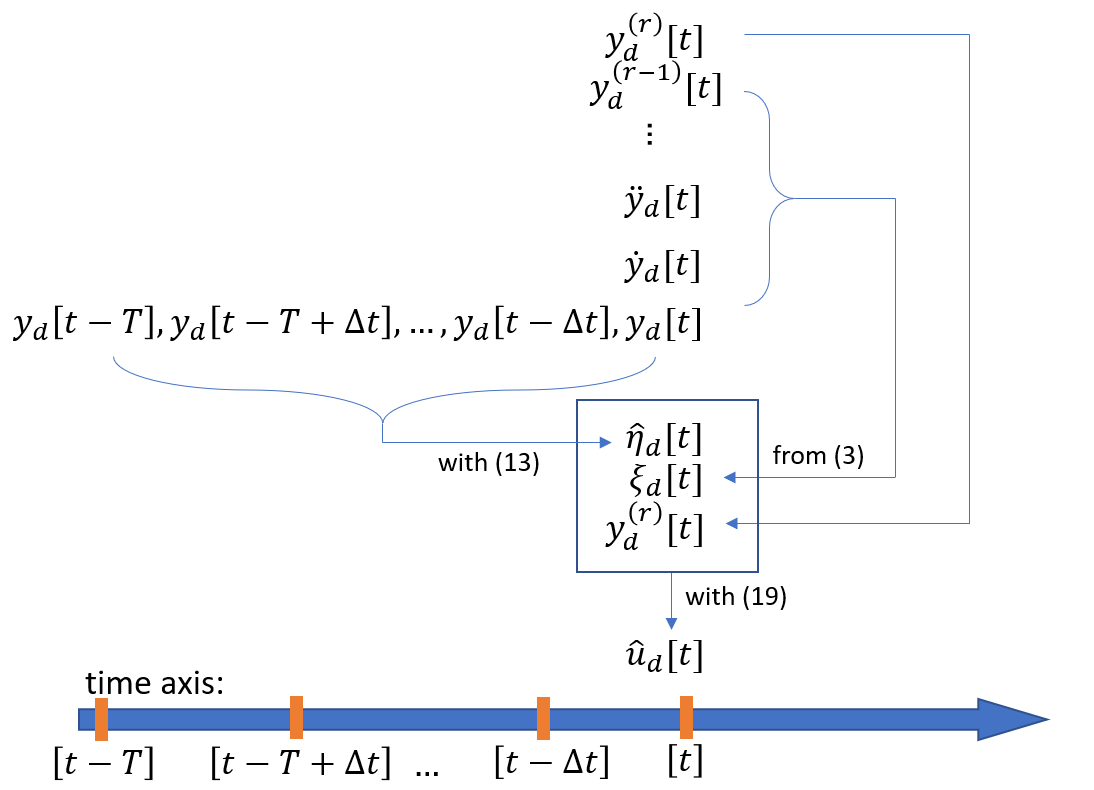}
\caption{The inverse operator's dependence on the  hidden state is removed by  use of past output history
and current time derivatives of the output. }
\label{fig_inverse_hidden_state_depend_demo}
\end{figure}

\vspace{0.1in}
\begin{remark}
\label{rem_inverse_preicison}
In addition to sufficient time history (large $T$)  of the output to accurately  find the internal state (to let $\Delta \eta \longrightarrow 0$), information about the derivatives of the output (upto the relative degree $r$ at time $t$, i.e., $y_d^{(r)}(t), \xi(t)$) are also needed  for precisely computing  the inverse 
input $u_d$
in (\ref{op_inverse_min_phase})
as illustrated in Fig.~\ref{fig_inverse_hidden_state_depend_demo}.
\end{remark}

\vspace{-0.1in} 

\vspace{-0.01in} 
\subsubsection{Koopman-type forward operators using output history\newline}
The output $y$ can be related to the input as 
\begin{equation}
\begin{split}
{y}(t+T_f) &=C\int_{-\infty}^{t+T_f}e^{A(t-\tau)}Bu(\tau)d\tau 
\end{split}    
\end{equation}
and approximated by 
\begin{equation}
\begin{split}
\hat{y}(t+T_f) &=C\int_{t-T}^{t+T_f}e^{A(t-\tau)}Bu(\tau)d\tau .
\end{split}   
\label{forward_map_approx_u_hist}
\end{equation}

Therefore, using arguments similar to the proof of Lemma 1, the error in computing the output using just the history of input $u$ tends to zero as the 
time history
of the input 
increases, i.e., as $T \rightarrow \infty$. Thus, it is possible to find a map that only depends on the input and its past history, 
\begin{equation}
\begin{split}
\hat{y}(t+T_f) & = 
\hat{\mathbb{G}}_u[u(t-T:t+T_f)],
\end{split}    
\end{equation}
which justifies the use of ARX models to capture forward linear system models using past input history (and augmented by the output history).  In contrast, with Koopman-type operators where past history of the observable output is used to predict future values, 
the forward model prediction can be written as 
\begin{equation}
\begin{split}
&\hat{y}(t+T_f)\\
&=Ce^{AT_f}\hat{x}(t)+C\int_{t}^{t+T_f}e^{A(t+T_f-\tau)}Bu(\tau)d\tau\\
&=Ce^{AT_f} S^{-1} \begin{bmatrix}
\xi(t)\\ \hat{\eta}(t)
\end{bmatrix}  +C\int_{t}^{t+T_f}
\!\!\! e^{A(t+T_f-\tau)}Bu(\tau)d\tau\\
& \qquad {\mbox{using (\ref{eq:coord_trans})} } \\
&=Ce^{AT_f} S^{-1} \begin{bmatrix}
\xi(t)\\ \hat{\mathbb{H}}[y_d(t-T:t)](t)
\end{bmatrix}  \\
& \qquad \qquad \qquad \qquad +C\int_{t}^{t+T_f}
\!\!\! e^{A(t+T_f-\tau)}Bu(\tau)d\tau\\
& \qquad {\mbox{using (\ref{eq_approx_unknown_state})} } \\
&\triangleq \hat{\mathbb{G}}[y(t-T:t), \xi(t), u(t:t+T_f)].
\end{split}    
\end{equation}
Therefore, past history of the output can also be used to develop Koopman-type forward operators, 
provided access is available to current time derivatives of the output $\xi(t)$.

\subsubsection{Inverse operator precision}
The inverse operator depends not only on the   past history of the output (to remove the hidden state $\eta$ dependency)
but also on the output and its time derivatives at the current time instant $t$. The impact of the time 
history $T$,
output and its time derivatives on the precision of the 
 operator 
is quantified in the next lemma. 

\begin{lemma}
\label{Lemma_prediction_error}
The prediction error of the inverse operator is bounded, i.e
there exists positive scalars $L_1>0, L_2>0, L_3 >0$  such that 
the error between the predicted input  $\hat{u}_d(t)$ and the true input $u_d(t)$ 
is
\begin{equation}
\begin{split}
&\|\hat{u}_d(t)-u_d(t)\|_{2}\\
&\le L_1\|\Delta y^{(r)}_d(t)\|_2 + L_2\|\Delta \xi_d(t)\|_2+L_3
e^{-\alpha_1 T}.
\end{split}
\label{eq:inv_u_err}
\end{equation}

\end{lemma}
\begin{pf}
From (\ref{eq:inv_model}) and (\ref{eq:inv_op_derivation}),
\begin{equation}
\begin{split}
&\|\hat{u}_d(t)-u_{d}(t)\|_{2}\\
&\le | b_{n-r}^{-1}|
\left[\|\Delta y^{(r)}_d(t)\|_2+\|A_{\xi}\|_2\|\Delta \xi_d(t)\|_2 \right. \\
& \qquad \qquad \qquad \qquad\left. +
\|A_{\eta}\|_2\|\Delta \eta_d(t)\|_2\right], 
\end{split}
\label{eq:quasi_err_step1}
\end{equation}
where $\Delta y^{(r)}_d(t) \triangleq \hat{y}^{(r)}_d(t)-y^{(r)}_d(t)$, $\Delta \xi_d(t)\triangleq \hat{\xi}_d(t)-\xi_d(t)$ and $\Delta \eta_d(t)\triangleq \hat{\eta}_d(t)-\eta_d(t)$.
The results follows from (\ref{eq_eta_err_bound_exp}) with 
\begin{equation}
L_1 = |b^{-1}_{n-r}|, \quad L_2 = L_1\|A_{\xi}\|_2,  \quad 
L_3 = L_1\|A_{\eta}\|_2   \beta_1 .
\end{equation}
\end{pf}

\vspace{0.1in}
\begin{remark}[Data-enabled algorithm]
\label{rem_Data_based_algorithm}
Known values of the desired output and its derivatives, specified with a sampling period $\Delta t$  and time history $T$ can be used to estimate a discrete-time inverse 
operator 
from (\ref{eq:inv_op_derivation}) as
\begin{align}
\hat{u}_d[m] &= \mathbb{G}_d^{-1}[y_d[m-m_T: 1:m],\xi_d[m],y^{(r)}_d[m]],
\label{eq_data_inverse}
\end{align}
where $[m]$ indicates value at time $t_m=m \Delta t$, and $m_T = T/{\Delta t} $.
Data-enabled algorithms can be used to learn the operator $\mathbb{G}_d^{-1}$, since  (\ref{eq_data_inverse})
maps a finite number of variables (desired output and its time  derivatives) to the inverse input at time $t_m$. 
\end{remark}

\section{Simulation results}
In this section, an example system is introduced, followed by the data-enabled learning of the inverse operator.

\subsection{Example system}
\begin{figure}[!ht]
  \centering
  \includegraphics[width=0.75\columnwidth]{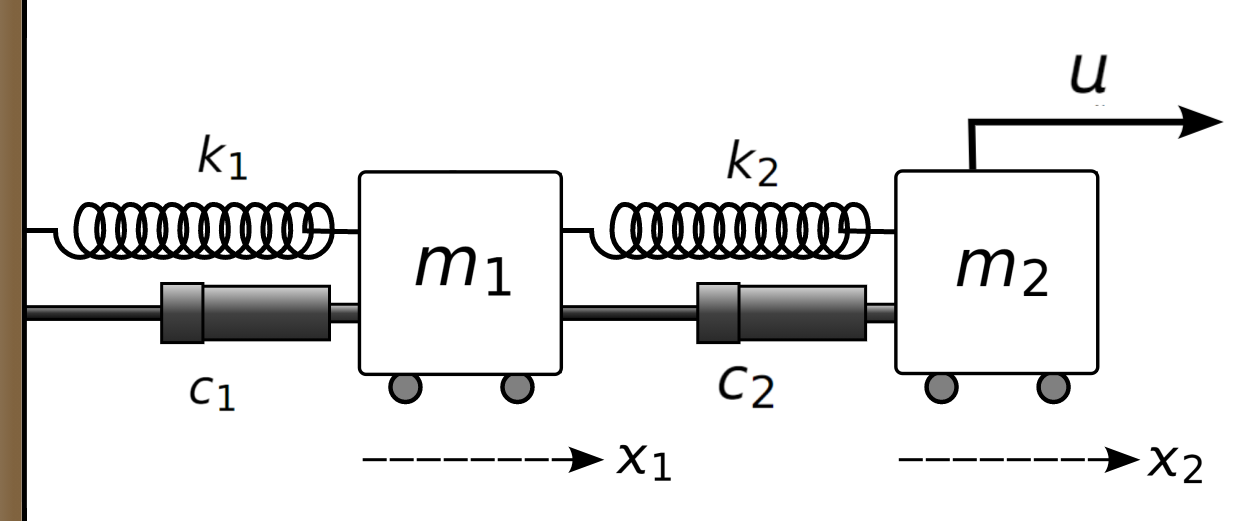}
  \caption{Example system plot}
  \label{fig:example_sys}
\end{figure}

Consider the following two-mass-spring-damper system, where the input $u$ is the force acting on mass $m_2$ and its displacement $x_2$ is the output $y$, as shown in Fig.~\ref{fig:example_sys}.

The corresponding state space model can be written as
\begin{align}
\frac{d}{dt} X &= AX+Bu\\    
y=x_2&=CX
\end{align}
where $X\triangleq \begin{bmatrix}
x_1&\Dot{x}_1&x_2&\Dot{x}_2
\end{bmatrix}'$,
$C=\begin{bmatrix}
0&0&1&0
\end{bmatrix}$, 
\begin{equation}
A=\begin{bmatrix}
0&1&0&0\\
-\frac{k_1+k_2}{m_1}&-\frac{c_1+c_2}{m_1}&\frac{k_2}{m_1}&\frac{c_2}{m_1}\\
0&0&0&1\\
\frac{k_2}{m_2}&\frac{c_2}{m_2}&-\frac{k_2}{m_2}&-\frac{c_2}{m_2}
\end{bmatrix}
, B=
\begin{bmatrix}
0\\0\\0\\ a/m_2
\end{bmatrix}, 
\end{equation}
$m_1=10,m_2=5,k_1=110,c_1=68,a=k_1/2,k_2 = 75$ and $c_2=60$ in SI units. The relative degree of the system is $r=2$ and the input-output relation is given by 
\begin{equation}
\begin{split}
\ddot{y}(t)&=
-25 y(t) -12 \dot{y}(t) +25 x_1(t) +12 \dot{x}_1(t)
+ {11 u(t)}.
\end{split}
\label{eq_yddot_example}
\end{equation}

\subsection{Preliminary selections}\label{subsec_pre}
\begin{figure}[!ht]
\centering
 \begin{tabular}{@{}cc@{}}
\includegraphics[width=0.44\columnwidth]{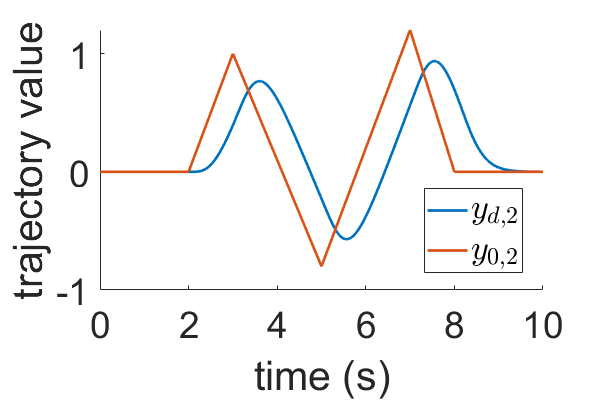} 
& 
\includegraphics[width=0.45\columnwidth]{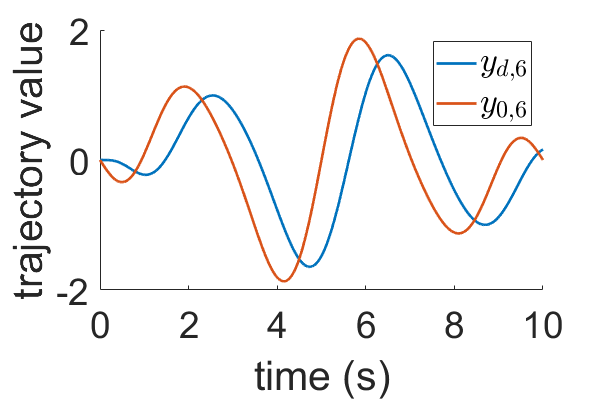}
\end{tabular}
  \vspace{-0.1in}
\caption{Comparison of the example filtered desired output $y_{d,k}$ and nominal trajectories $y_{0,k}$  for $k=2$ (triangular) and $k=6$ (sinusoidal). 
}
\label{fig_test_traj_demo2}
\end{figure}

Selection of the data-enabled model types to evaluate, the sampling time (which needs to be sufficiently small to reduce discretization error), the evaluation metric, and sufficiently  smooth output trajectories for model  evaluation are described below. 
\vspace{-0.1in}
\begin{enumerate}[label=(\roman*)]   
\item A two-layer feedforward neural-net (created through MATLAB function \texttt{feedforwardnet()} with default activation function) is used to learn the inverse operator 
from data. 
\item For the two-layer neural-net, each model pool consists of 5 candidates with different
 number 
$N\in\{5,10,20,40,80\}$ of neurons in the hidden layer
\item 
The sampling frequency is varied from $5$ Hz to $20$ Hz, which is substantially higher than the system bandwidth of 1.7 Hz.

\begin{figure}[!ht]
\centering
\includegraphics[width=0.95\columnwidth]{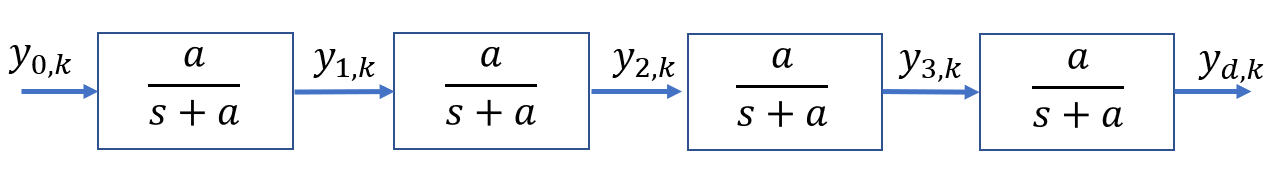}
\caption{Filter process to generate desired trajectories. 
}
\label{fig:sig_filter}
\end{figure}
\item The inverse operator is assessed using  10 different desired trajectories $y_{d,k}(t), 1\le 
k
\le 10, t\in [0,10]$ with a fixed prediction sampling time of $0.01$ s.
Each desired  trajectory $y_{d,k}$ 
used for assessment needs to be sufficiently smooth to investigate the impact of different order of output's time derivatives  on the inverse operator, although from  (\ref{eq:inv_u_err}) the expectation is that only output derivatives  upto the $r^{th}$ order ($r=2$ for this example) are required. Therefore, nominal trajectories $y_{0,k}$ (specified in the appendix) are filtered as shown in Fig.~\ref{fig:sig_filter}, to obtain desired outputs $y_{d,k}$ and their derivatives as 
 \begin{equation}
\begin{bmatrix}
 y_{d,k}\\\Dot{y}_{d,k}\\\Ddot{y}_{d,k}\\y^{(3)}_{d,k}\\y^{(4)}_{d,k}
\end{bmatrix}(t)
=
\begin{bmatrix}
1&0&0&0&0\\
-a&a&0&0&0\\
a^2&-2a^2&a^2&0&0\\
-a^3&3a^3&-3a^3&a^3&0\\
a^4&-4a^4&6a^4&-4a^4&a^4
\end{bmatrix}
\begin{bmatrix}
y_{d,k}\\y_{3,k}\\y_{2,k}\\y_{1,k}\\y_{0,k}
\end{bmatrix}(t)
\label{eq_filter_compu}
\end{equation}
where $a=2\pi$ (cut-off frequency as 1 Hz), which is less than the system's bandwidth of 1.7 Hz, and example trajectories are shown in Fig.~\ref{fig_test_traj_demo2}. 
\item 
For a given time history $T$ and  sampling time $\Delta t$,  as in Remark~\ref{rem_Data_based_algorithm}, the evaluation metrics for the data-enabled inverse  operator with $N$ neurons in the hidden layer are selected as the mean $e_{u,N}$ and maximum $\overline{e}_{u,N}$
normalized prediction error 
over the ten 
evaluation trajectories 
$y_{d,k}(\cdot)$, i.e., 
\begin{equation}
e_{u,N} = \frac{1}{10}\sum_{k=1}^{10}\frac{\max_{m} |\hat{u}_k[m]-u_{d,k}[m]|}{\max_{m} |u_{d,k}[m]|}\times 100\%
\label{metric_1}
\end{equation}
\begin{equation}
\overline{e}_{u,N} = \max_{k=1,\dots,10}\frac{\max_{m} |\hat{u}_k[m]-u_{d,k}[m]|}{\max_{m} |u_{d,k}[m]|}\times 100\%,
\label{metric_2}
\end{equation}
where
the ideal inverse $u_{d,k}$ was found 
using (\ref{eq:inv_model}) where $\eta_d$ was obtained through (\ref{eq:unknown_state_by_integral}).
Moreover, the smallest normalized prediction error over different number of neurons in the hidden layer is defined as 
\begin{equation}
e_{u} =  e_{u, N^*}, \quad 
\bar{e}_{u} =  \bar{e}_{u, N^*} \quad 
{\mbox{where} }\quad
N^* = \argmin_N{e_{u, N}}
\label{metric_3}
\end{equation}
to quantify the precision of the   inverse operator.
\end{enumerate}

  \vspace{-0.1in}
  \subsection{Data Collection }
  \vspace{-0.1in}
The inverse operators
are trained using input-output data collected from simulations. Both noisy and noise free output data are used to assess the impact of noise. 
The input signal $u$ applied to the system is constructed by concatenating $20$ cycles of $p_{(f_i,\alpha_i)}(\cdot)$ ($i=1,2,3,\dots,20$) with different parameters, which are tabulated in Table.~\ref{tab:excitation_sig}.
\begin{equation}
p_{(f_i,\alpha_i)}(t)=\alpha_i [ 4\sin{(\pi c t^2)}+s(t)+r(t) ]
\label{eq:excitation_sig}
\end{equation}
where $c = f_i/10$,
\begin{equation*}
\text{s}(t)
=\begin{cases}
1 & 2\le t < 4\\
-0.9 & 4\le t < 6\\
0.5 & 6\le t < 8\\
0 & \text{otherwise},
\end{cases}
\quad
\text{r}(t)=\begin{cases}
0.4t & 0\le t < 1\\
0.4 & 1\le t < 9\\
\text{r}(10-t) & 9\le t \le 10.
\end{cases}
\end{equation*}

\begin{table}[!ht]
\centering
\caption{Parameters of  $p_{(f_i,\alpha_i)}$ in Eq.~\eqref{eq:excitation_sig}. 
}
\begin{tabular}{|c|c|c|c|c|c|}
\hline
Cycle $\#$, $i$ & 
$f_i$ & $\alpha_i$ &
Cycle $\#$, $i$ & 
$f_i$ & $\alpha_i$\\
\hline
1   & 6 & 0.75 &11   & 1 & 0.25  \\
2   & 3 & 0.5 &12   & 0.5 & 0.25\\
3  & 2 & 0.5 &13   & 1 & -0.1\\
4  & 0.5 & 0.5 &14   & 0.5 & -0.05\\
5 & 0.5 & 0.3 &15   & 0.5 & 0.1\\
6  & 0.3 & 0.3 &16  & 0.5 & -0.1\\
7  & 0.1 & 0.3 &17   & 2 & 0.25\\
8  & 0.5 & -0.3 &18   & 1 & 0.1\\
9  & 0.3 & -0.3 &19   & 0.5 & 0.05\\
10  & 0.1 & -0.3 &20   & 1 & 0.5\\
\hline
\end{tabular}
\label{tab:excitation_sig}
\end{table}
For the noisy case,  additive white gaussian noise with signal-to-noise ratio of 20 is separately added to each output and its time derivatives.
Simulations were  done in MATLAB with \texttt{ode45()} with sampling rate of 100 Hz (to be consistent with the evaluation metrics 
from (\ref{metric_1}) to (\ref{metric_3})).
 Input, output and the output's time derivatives (upto the fourth order) were collected. Second order derivative 
was
 obtained from~(\ref{eq_yddot_example}). Third and fourth order  derivatives for training purposes were estimated from the data, using finite difference as, 
\begin{align}
\begin{bmatrix} y^{(3)}[m]\\
y^{(4)}[m]
\end{bmatrix} 
&=\frac{1}{12(\Delta t)}\begin{bmatrix}
-1&8&0&-8&1 \\
-\frac{1}{\Delta t}&\frac{16}{\Delta t}&-\frac{30}{\Delta t}&\frac{16}{\Delta t}&-\frac{1}{\Delta t}
\end{bmatrix}\begin{bmatrix}
\ddot{y}[m+2]\\\ddot{y}[m+1]\\\ddot{y}[m]\\\ddot{y}[m-1]\\\ddot{y}[m-2]
\end{bmatrix}.
\nonumber 
\end{align}

\begin{figure}[!t]
  \centering
  \includegraphics[width=0.75\columnwidth]{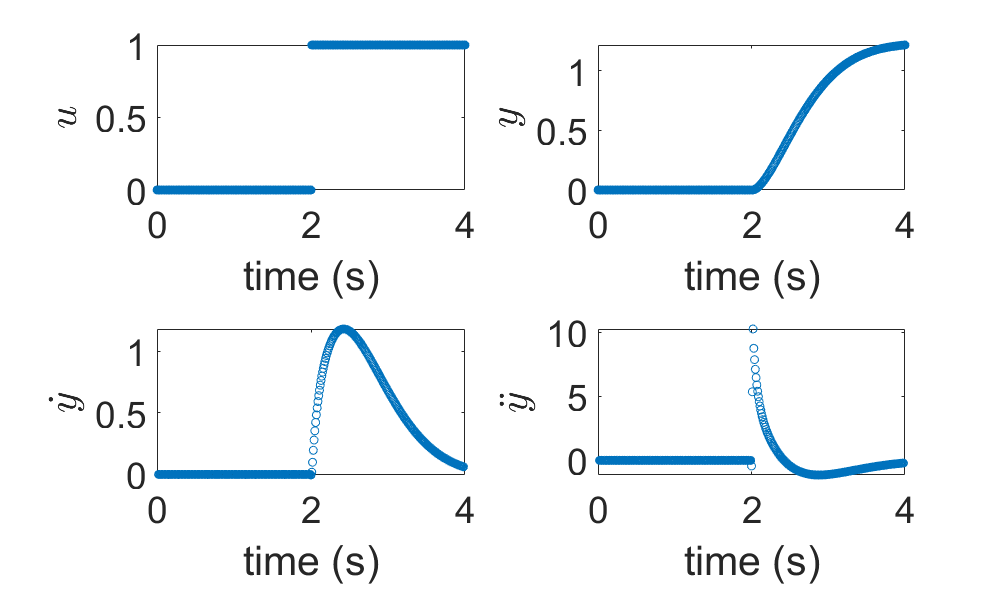}
  \vspace{-0.1in}
  \caption{Identifying the relative degree $r$ from input-output data, based on discontinuity in the $r^{th}$ derivative of the output for a step input. 
  }
  \label{fig_relative_degree_id}
\end{figure}
\vspace{0.1in}

\subsection{Reducing impact of hidden states using output  history }
\begin{figure}[!t]
\centering
\includegraphics[width=0.75\columnwidth]{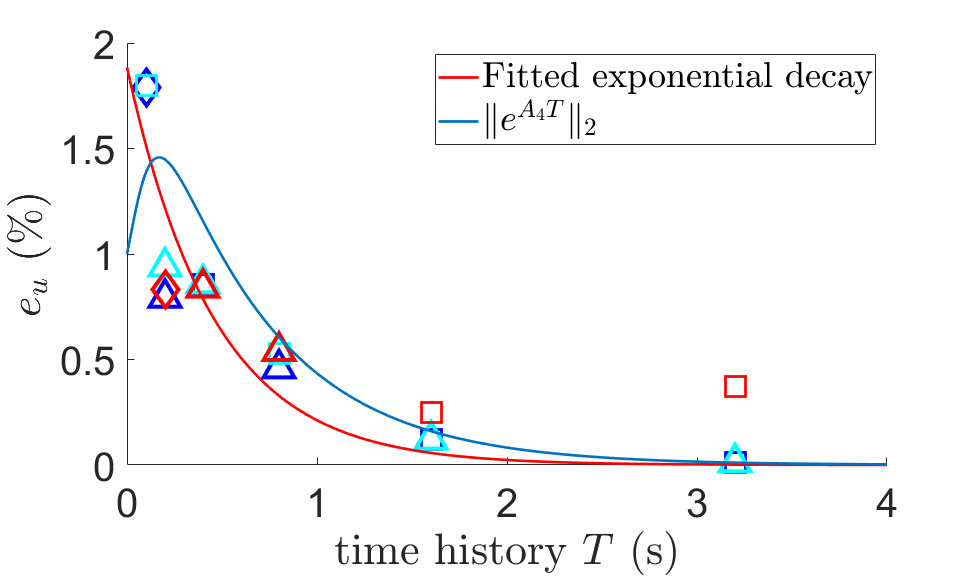}
\caption{Inverse 
operator's 
precision in terms of prediction error  $e_u$  (\ref{metric_3})  exponentially improves with respect to different window length $T$ of output history,  for different sampling times,  $\Delta t = 0.05 s(20 \text{Hz, blue}), 0.1 s(10 \text{Hz, cyan}), 0.2 s(5\text{Hz, red})$.
Similar results are seen over different 
$N^*$ neurons in the hidden layer:
$5$  triangle ($\triangle$), $10$ (square $\square$),$20$ (diamond $\diamondsuit$),$40$ (pentagram $\medwhitestar$), and $80$ (circle $\fullmoon$). The fitted exponential decay (red line) is obtained with sampling time of $\Delta t = 0.05$  s ($20$ \text{Hz, blue}).}
\label{fig_inv_exponential_decay}
\end{figure}

To investigate the reduction of the impact of  the hidden states on the prediction precision of data-enabled inverse operators, 
the performance of the data-enabled inverse operators
was assessed for different 
time history
$T$ of the output. In this part of the study,  the number of time derivatives of the output used was the same as the relative degree of the example system.
The relative degree $r=2$  can be established by applying a step input ---  a corresponding discontinuity will appear in ${y}^{(r)}$, while the lower order derivatives   ($y,\dot{y}$ in this example) remain continuous as seen in Fig.~\ref{fig_relative_degree_id}.
Then, from (\ref{eq_data_inverse}), 
\begin{align}
\hat{u}_d[m] &= \mathbb{G}^{-1}_d[y_d[m-m_T:1:m],\Dot{y}_d[m],\Ddot{y}_d[m]].
\label{eq:experiment_data_inverse}
\end{align}
The inverse operator's 
prediction error  $e_u$ (\ref{metric_3}) was obtained for varying output 
time history
$T$ ([0.1, 0.2, 0.4, 0.8, 1.6, 3.2, $ \dots$] s), for different sampling time $\Delta t \in \left\{ 0,05 s, 0.1 s, 0.2 s \right\} $, and  for different number 
$N$ of neurons in the hidden layer, and  plotted in   Fig.~\ref{fig_inv_exponential_decay}  for the case without noise in the training data. The associated prediction errors are tabulated in  Table~\ref{tab_excitation_sig_mean}  for the fastest sampling time $\Delta t = 0.05$ s. 

The precision of the inverse operator
improves with larger output time history 
$T$, as seen in Table~\ref{tab_excitation_sig_mean},  where the evaluation values of the two-layer neural net with
different $N$ neurons in the hidden layer are listed.
Note that typically  $N^*\le 20$ yields good precision for this application from Table~\ref{tab_excitation_sig_mean}.
Over all selections of neuron numbers $N$, the variation of the smallest prediction error $e_u=e_{u,N^*}$ 
(\ref{metric_3}) with sampling time of $\Delta t = 0.05$  s ($20$ Hz)   fits an exponential decay curve  $e_u(T) \approx 1.88e^{-2.18T}$, shown in red in Fig.~\ref{fig_inv_exponential_decay}.
This exponential improvement in precision  is 
expected from Lemma~\ref{Lemma_prediction_error}, 
which predicts an exponential decay of error in the estimation of the hidden states, dependent on  $\|e^{AT}\|_2$ from (\ref{eq:exponeital_decay}), and  shown in Fig.~\ref{fig_inv_exponential_decay}.
Thus, the impact of hidden states on the prediction precision of data-enabled inverse operator
can be reduced by using sufficient time history of the desired output.

\vspace{0.1in}
\begin{remark}[Reducing hidden state dependence]
In the following simulations, the time history $T$ is chosen to be sufficiently large $T^*=3.2$ s, which results in a 
normalized 
error $e_u \approx 0.01\%$. 
\label{remark_select_Tstart}
\end{remark} 

\vspace{0.1in}
\begin{table}[!t]
\centering
\caption{Inverse 
operator's 
precision improvement in terms of prediction  error $e_{u,N}$ (\ref{metric_1}) and $\overline{e}_{u,N}$ (\ref{metric_2})  for varying output time  history  $T$ and number $N$ of neurons in the hidden layer, with sampling time $\Delta t = 0.05$ s.}
\begin{tabular}{|c|c|c|c|c|c|}
\hline
\diagbox{T}{N} & 5 & 10 & 20 & 40 & 80 \\
\hline
&  \multicolumn{4}{c}{$e_{u,N} (\%)$ as in (\ref{metric_1})} & \\
\hline
0.1&1.78 &2.23 & 1.64 & 2.05 & 2.43\\
0.2&0.79 &0.88 & 0.87 & 0.88 & 0.98\\
0.4& 0.95& 0.85& 0.88 & 0.92 & 0.91\\
0.8& 0.46&0.51 & 0.49 & 0.48 &0.52 \\
1.6& 0.14&0.12 & 0.12 & 0.14 & 0.16\\
3.2&0.05 & 0.01 & 0.01 & 0.01 & 0.05\\
\hline
&  \multicolumn{4}{c}{$\overline{e}_{u,N}(\%)$ as in (\ref{metric_2}) } & \\
\hline
0.1&3.17 & 3.72& 4.73 & 5.61 & 6.28\\
0.2& 1.22& 1.59& 1.56 & 1.48 & 1.76 \\
0.4& 1.17&1.10 & 1.33 & 1.75 &  1.69\\
0.8&0.54 &0.65 &0.61  & 0.67 &1.01  \\
1.6&0.20 & 0.16& 0.18 & 0.33& 0.44 \\
3.2& 0.08&0.02 &0.02 &  0.02& 0.14 \\
\hline
\end{tabular}
\label{tab_excitation_sig_mean}
\end{table}   

  \vspace{-0.1in}
  \subsection{Need to include output time derivatives}
  \vspace{-0.1in}
From (\ref{eq:inv_u_err}) in Lemma~\ref{Lemma_prediction_error}, 
even if the hidden state error is reduced by having sufficiently large time history $T$, (as shown in the previous subsection), current time derivatives of the output  $\xi_d(t),y^{(r)}(t)$ are needed to achieve precision prediction with the inverse 
operator. Therefore, the impact of adding time-derivative information is investigated through the following two steps, for different sampling periods $\Delta t \in \left\{ 0,05 s, 0.1 s, 0.2 s \right\}$ and  for different number $N$ of neurons in the hidden layer. 

\begin{enumerate}[label=(\roman*)]
\item Incrementally including higher-order  time derivatives of the output when learning the inverse operator $\mathbb{G}^{-1}_{d,l}$ that predicts 
the inverse input $\hat{u}_d$ 
similar to  (\ref{eq:experiment_data_inverse}),  where  output time derivatives till order $l$  ($0 \le l \le 4$) 
are included in the data-enabled 
operator learning, e.g., with $l=i \ge 0$,
\begin{align}
\hat{u}_d[m] & =
\mathbb{G}^{-1}_{d,i}[y_d[m-m_T:1:m],
\nonumber \\
& \qquad \quad  y^{(i)}_d[m], y^{(i-1)}_d[m], \hdots  y^{(0)}_d[m]],
\label{inv_G_d_4}
\end{align}
where $\mathbb{G}^{-1}_{d,2} = \mathbb{G}^{-1}_{d}$ in (\ref{eq:experiment_data_inverse}).

 \item  Adding the output's time derivatives $\dot{y}_d(t),\ddot{y}_d(t)$ to NARX-type 
 inverse operators where the inverse operator is learned using both input and output time history, i.e., to compare
 \begin{equation}
\begin{split}
\hat{u}_d[m] = &  \text{NARX}[y_d[m-m_T:1:m], \\
& \quad u_d[m-m_T:1:m-1]]
\end{split}
\label{eq_narx}
\end{equation}
\begin{equation}
\begin{split}
\hat{u}_d[m] = &  \text{NARX}^{*}[y_d[m-m_T:1:m],\dot{y}_d[m],
\\
& \quad  \ddot{y}_d[m], u_d[m-m_T:1:m-1]]. 
\end{split}
\label{eq_narx_star}
\end{equation}
\end{enumerate}

\begin{figure}[!ht]
\centering
 \begin{tabular}{@{}c@{}}
 \includegraphics[width=0.9\columnwidth]{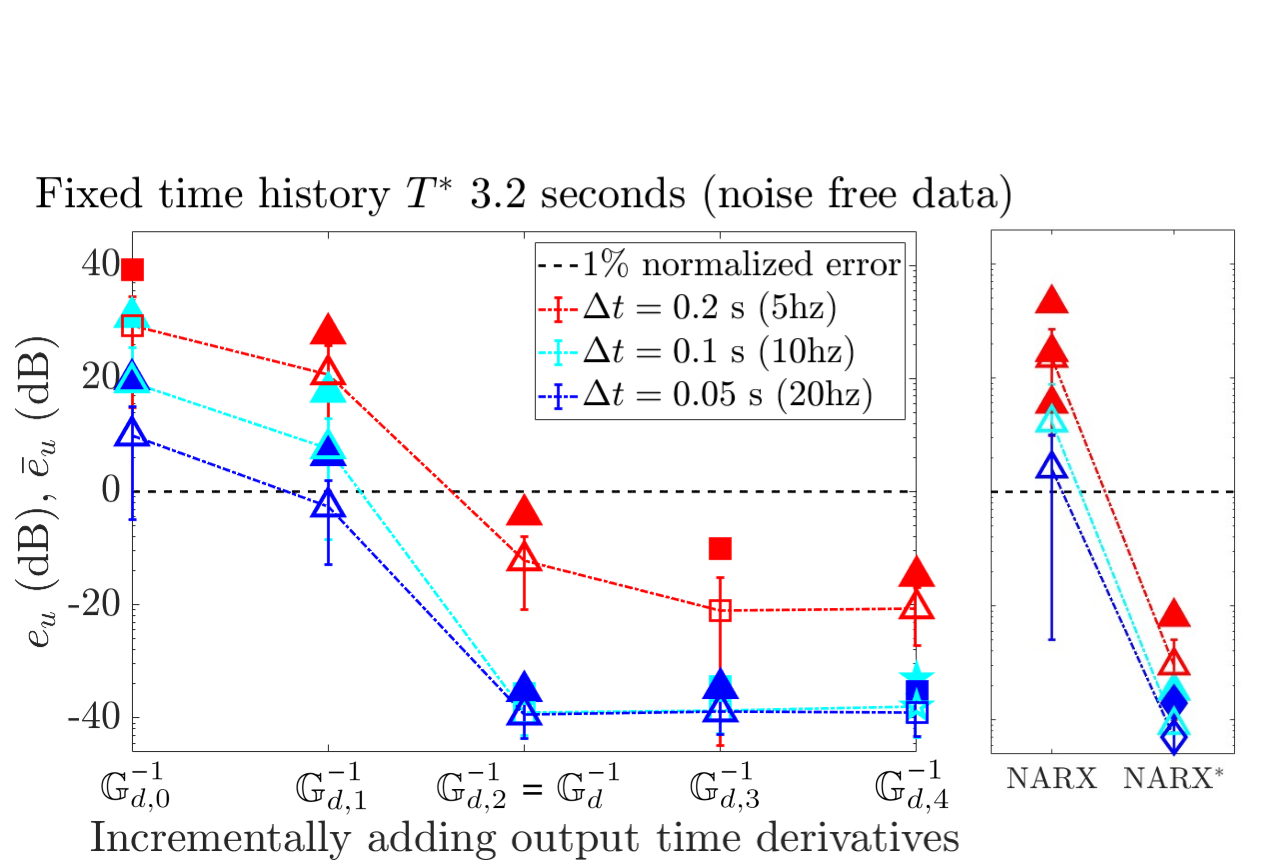}\\
\includegraphics[width=0.9\columnwidth]{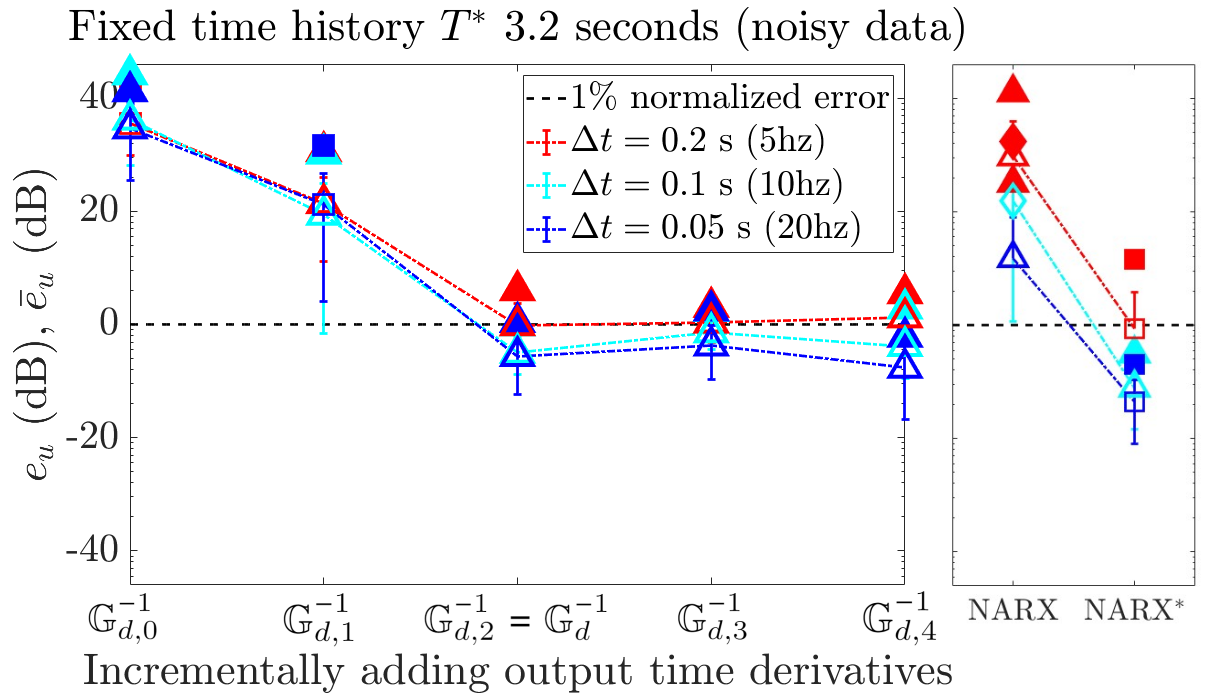}
  \end{tabular}
\caption{
Inverse operator's 
precision in terms of prediction error  $e_u, \overline{e}_u$  (\ref{metric_3}) improves for all cases with the addition of derivative information.
(Top: noise free training data. Bottom: noisy training data).
Similar results are seen for different 
number $N^*$ of neurons in the hidden layer,
with symbols as in Fig.~\ref{fig_inv_exponential_decay}, where the filled symbols correspond to  $\overline{e}_u$ and unfilled correspond to ${e}_u$.
Performance of  NARX-type 
operator 
with input and output history but without  derivative information is also improved with the addition of derivative information in  $\text{NARX}^*$, as in  (\ref{eq_narx},\ref{eq_narx_star})}
\label{fig_inv_map_noise_free}
\end{figure}

The corresponding prediction performance, in terms of errors $e_u$ and $\bar{e}_u$ in (\ref{metric_3}), for $T^*=3.2$ s and $\Delta t = 0.05$~s
are tabulated in Table~\ref{tab_derivative_impact}, and plotted in Fig~\ref{fig_inv_map_noise_free} for $T^*=3.2$~s and different sampling time $\Delta t \in \{0.05s, 0.1s, 0.2s\}$.

\begin{table}[!ht]
\centering
\caption{Prediction error $e_u,\bar{e}_u$ (\ref{metric_3}) for inverse 
operators
from (\ref{inv_G_d_4}) to (\ref{eq_narx_star}) 
with $\Delta t = 0.05$ s.}
\begin{tabular}{|c|c|c|c|c|c|}
\hline
 & $e_u(\%)$ & $\bar{e}_u (\%)$ & & $e_u(\%)$ & $\bar{e}_u(\%)$ \\
\hline
&  \multicolumn{4}{c}{Noise free training data} & \\
\hline
$\mathbb{G}^{-1}_{d,0}$& 3.13& 9.82 & $\mathbb{G}^{-1}_{d,4}$ & 0.01 & 0.02\\
$\mathbb{G}^{-1}_{d,1}$& 0.74& 2.10 & NARX & 1.60 & 5.93\\
$\mathbb{G}^{-1}_{d,2} =\mathbb{G}^{-1}_{d}$&  0.01& 0.02 & $\text{NARX}^*$ & 0.01 & 0.02\\
$\mathbb{G}^{-1}_{d,3}$& 0.01& 0.02&  &  & \\
\hline
&  \multicolumn{4}{c}{Noisy training data} & \\
\hline
$\mathbb{G}^{-1}_{d,0}$& 53.91 & 114.68 & $\mathbb{G}^{-1}_{d,4}$ & 0.41 & 0.78\\
$\mathbb{G}^{-1}_{d,1}$& 11.53& 37.82& NARX & 3.89 & 17.95\\
$\mathbb{G}^{-1}_{d,2} =\mathbb{G}^{-1}_{d}$   &  0.53 & 1.05 & $\text{NARX}^*$ & 0.21 &0.45 \\
$\mathbb{G}^{-1}_{d,3}$& 0.65& 1.32&  &  & \\
\hline
\end{tabular}
\label{tab_derivative_impact}
\end{table}   

{\underline{Impact of including derivatives}}
The precision of the inverse 
operator
depends on inclusion of the  output derivative upto order $r$ (the relative degree). When the number of derivatives $l$  (included in the training and evaluation) is increased 
from $l=0$ to $l=4$, the precision of the inverse 
operator
improves significantly when all the required number ($l=2=r$) of  time derivative features are included in the training and evaluation data. In particular, the maximum error $\overline{e}_u$ in  (\ref{metric_3}) reduces from $9.82\% $ to $0.02\%$  for the case with noise free training data and 
from $114.68\% $ to $1.05\%$  for the case with noisy training data as seen in Table~\ref{tab_derivative_impact}. Therefore, there is substantial improvement in the inverse 
operator's 
precision (especially in the presence of noise) when  time derivatives upto the required order of 2 are included.

{\underline{Impact on NARX-type inverse operator}}
Inclusion of time derivatives is also important for NARX-type inverse
operators
where both input and output time history 
are used in the inverse 
operator. This can be seen by comparing NARX (\ref{eq_narx}) without time derivatives  and $\text{NARX}^*$ (\ref{eq_narx_star}) with the derivatives in  Table~\ref{tab_derivative_impact} and in Fig~\ref{fig_inv_map_noise_free}.
When time derivatives $l=2$  are included in the training and evaluation,  the precision of the inverse
operator
improves significantly. 
In particular, the maximum error $\overline{e}_u$ in  (\ref{metric_3}) reduces from $5.93\% $ to $0.02\%$  for the case with noise free training data and 
from $17.95\% $ to $0.45\%$  for the case with noisy training data
as seen in Table~\ref{tab_derivative_impact} .
Therefore, there is substantial improvement in the precision of the NARX-type inverse
operator
when the output time derivatives upto the required order of 2 are included.

{\underline{Derivative information in output time history}}
Conceptually, information about the derivatives upto $r-1$ (one less than the relative degree $r$) are available in the time history of the output and only the $r^{th}$ time derivative $y_d^{(r)}[m]$  is directly affected by the input $u[m]$. 
In particular, output derivatives can  be related to the output time history  using finite difference techniques, especially in the noise free case, and hence direct computation of the derivatives might not appear to be critical if time history of the output is used  during training. Nevertheless, including computed or  measured  values (even with some noise) of the time derivative $\dot{y}[m]$ (which is not directly affected by the input $u[m]$) 
still can improve
the precision of the inverse 
operator
as seen in Fig.~\ref{fig_inv_map_noise_free} and Table~\ref{tab_derivative_impact}. 
In particular, the maximum error $\overline{e}_u$ in  (\ref{metric_3}) reduces from $9.82\% $ to $2.10\%$  for the case with noise free training data and 
from $114.68\% $ to $37.82\%$  for the case with noisy training data as seen in Table~\ref{tab_derivative_impact}. Therefore, while the noise free case precision could be improved by smaller sampling time $\Delta t$ without the inclusion of $\dot{y}$, for the noisy case,  direct measurements of the output time derivatives can substantially improve the inverse operator training, and lead to better precision in its predictions. Moreover, the precision of the inverse operator is further improved by including time derivatives upto the required order of r (relative degree). 

\section{Conclusion}

\vspace{-0.1in}
This work showed that Koopman-type data-enabled inverse operators  can have high precision if a sufficient large time history  of the output is included to reduce the impact of hidden internal states. Additionally,  measurements of the instantaneous output time derivatives (upto the relative degree) are required during training to improve the 
data-enabled inverse operator
precision. Our ongoing work is aimed at extending these results to Koopman-type data-enabled inverse operators for nonlinear nonminimum-phase systems.

\vspace{-0.1in}
\bibliography{paper}   

\appendix
\section{Evaluation trajectories}
Expressions of $y_{0,k}(t)$ for $k=1,2,\dots , 10$ and $0\le t \le 10$.
Trapezoidal shape ($k=1$)
\begin{equation*}
y_{0,1}(t)=\begin{cases}
0.4(t-1) & 1\le t <3\\
0.8 & 3\le t < 6\\
0.4(8-t) &6\le t <8\\
0&\text{otherwise}.
\end{cases}
\end{equation*}
Triangle wave ($k=2$)
\begin{equation*}
y_{0,2}(t)=\begin{cases}
t-2 & 2\le t <3\\
3.7-0.9t & 3\le t < 5\\
t-5.8 &5\le t <7\\
1.2(8-t)&7\le t <8\\
0&\text{otherwise}.
\end{cases}
\end{equation*}
Square wave ($k=3$)
\begin{equation*}
y_{0,3}(t)=\begin{cases}
1 & 2\le t <4\\
-1 & 4\le t < 6\\
1&6\le t <8\\
0&\text{otherwise}.
\end{cases}
\end{equation*}
Serrated wave mixture ($k=4$)
\begin{equation*}
y_{0,4}(t)=\begin{cases}
2(t-1)/3 & 1\le t <2.5\\
2(4-t)/3 & 2.5\le t < 4\\
8(t-4)/15 &4\le t <5\\
8(6-t)/15 &5\le t <6\\
0.4(t-6)&6\le t <7.5\\
0.4(9-t)&7.5\le t <9\\
0&\text{otherwise}.
\end{cases}
\end{equation*}
Monotonic ($k=5$): 
$y_{0,5}(t) = 0.001(x^{3.2}-x^2)$
Sine wave \#1 ($k=6$)
\begin{equation*}
y_{0,6}(t)=\sin(0.4\pi t)-0.9\sin(0.6\pi t)+0.2\sin(\pi t)
\end{equation*}
Sine wave \#2 ($k=7$)
\begin{equation*}
y_{0,7}(t)=1.5\sin(0.7\pi t)-0.5\sin(0.4\pi t)
\end{equation*}
Sine wave \#3 ($k=8$)
\begin{equation*}
y_{0,8}(t)=-0.5\sin(0.3\pi t)-0.6\sin(0.7\pi t)+0.2\sin (1.2\pi t)
\end{equation*}
Sine wave \#4 ($k=9$)
\begin{equation*}
y_{0,9}(t)=0.7\sin(0.26\pi t)+0.3\sin(1.3\pi t)-0.2\sin (1.4\pi t)
\end{equation*}
Slow chirp wave ($k=10$): 
$y_{0,10}(t)=0.35\sin (x^{1.5})$.

\end{document}